\documentclass[letterpaper]{article} 
\usepackage{aaai2026}  
\usepackage{times}  
\usepackage{helvet}  
\usepackage{courier}  
\usepackage[hyphens]{url}  
\usepackage{graphicx} 
\usepackage{natbib}  
\usepackage{caption} 
\usepackage{booktabs}
\usepackage{algorithm}
\usepackage{transparent}
\usepackage{amsmath}
\usepackage{amssymb}
\usepackage{graphicx}           
\usepackage{colortbl}
\usepackage{multirow}
\usepackage{algpseudocode}
\usepackage{color}
\usepackage{xcolor}
\usepackage{newfloat}
\usepackage{listings}
\DeclareCaptionStyle{ruled}{labelfont=normalfont,labelsep=colon,strut=off} 
\floatstyle{ruled}
\newfloat{listing}{tb}{lst}{}
\floatname{listing}{Listing}
\title{Vox-Evaluator: Enhancing Stability and Fidelity for Zero-shot TTS with A Multi-Level Evaluator}
\author{
    Hualei Wang\textsuperscript{\rm 1 }
    Na Li\textsuperscript{\rm 1*},
    Chuke Wang\textsuperscript{\rm 1},
    Shu Wu\textsuperscript{\rm 1},
    Zhifeng Li\textsuperscript{\rm 3}\thanks{Corresponding authors.},
    Dong Yu\textsuperscript{\rm 2},
}
\affiliations{
    \textsuperscript{\rm 1}Tencent AI Lab, ShenZhen
    \textsuperscript{\rm 2}Tencent AI Lab, Seattle
    \textsuperscript{\rm 3}Independent Author\\
    hualele.wang@gmail.com
}

\usepackage{bibentry}

\begin{document}

\maketitle

\begin{abstract}
Recent advances in zero-shot text-to-speech (TTS), driven by language models, diffusion models and masked generation, have achieved impressive naturalness in speech synthesis. Nevertheless, stability and fidelity remain key challenges, manifesting as mispronunciations, audible noise, and quality degradation. To address these issues, we introduce Vox-Evaluator, a multi-level evaluator designed to guide the correction of erroneous speech segments and preference alignment for TTS systems. It is capable of identifying the temporal boundaries of erroneous segments and providing a holistic quality assessment of the generated speech.
Specifically, to refine erroneous segments and enhance the robustness of the zero-shot TTS model, we propose to automatically identify acoustic errors with the evaluator, mask the erroneous segments, and finally regenerate speech conditioning on the correct portions. In addition, the fine-gained information obtained from Vox-Evaluator can guide the preference alignment for TTS model, thereby reducing the bad cases in speech synthesis. Due to the lack of suitable training datasets for the Vox-Evaluator, we also constructed a synthesized text-speech dataset annotated with fine-grained pronunciation errors or audio quality issues. The experimental results demonstrate the effectiveness of the proposed Vox-Evaluator in enhancing the stability and fidelity of TTS systems through the speech correction mechanism and preference optimization. 
The demos are shown.\footnote{\url{https://voxevaluator.github.io/correction/}}
\end{abstract}

\section{Introduction}
In recent years, zero-shot text-to-speech (TTS) achieves considerable advancements, particularly in synthesizing natural and expressive speech that is consistent in timbre and style with just a few seconds of audio prompt ~\cite{borsos2023audiolm,kharitonov2023speak,chen2025neural}. These models, built upon large language models (LLMs) ~\cite{zhang2023speak,chen2024vall,lajszczak2024base,du2024cosyvoice}, diffusion models ~\cite{jiang2025sparse,anastassiou2024seed,eskimez2024e2} and masked generative models ~\cite{wang2024maskgct,ju2024naturalspeech}, have demonstrated an extraordinary capability to generate speech with high naturalness, opening new frontiers for high-quality podcast generation ~\cite{ju2025mooncast} and audiobook production ~\cite{park2025multiactor}. In these application scenarios, stability and fidelity of TTS are the most important. However, producing stable and high-fidelity speech remains a significant challenge due to the randomness of the sampling process and the inherent difficulty in modeling complex long-form content.

\begin{figure}[!t]
    \centering
    \includegraphics[width=1.\columnwidth]{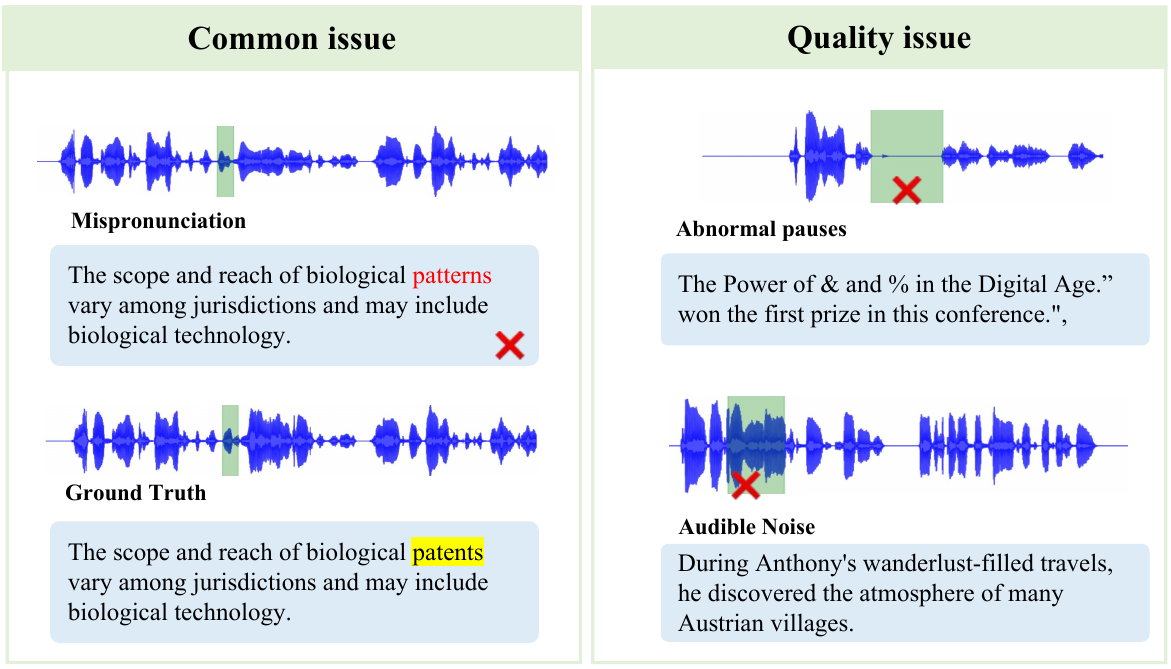}
    \caption{Typical instances of distinct erroneous samples. The error segments in the generated speech are emphasized by green bounding boxes.}
    \label{fig:error}
\end{figure}
Zero-shot TTS models are usually categorized into autoregressive (AR) and non-autoregressive (NAR) models. AR-based TTS models typically treat speech synthesis as a sequential prediction task and use LLM to autoregressively generate discrete tokens ~\cite{zhang2023speak,chen2024vall,lajszczak2024base,du2024cosyvoice}. Although AR models exhibit rich prosodic diversity, they inherently suffer from error accumulation and potential instability. NAR-based models, including those based on diffusion ~\cite{shen2023naturalspeech} and flow matching ~\cite{lipman2022flow} models, treat speech synthesis as a parallel generation and abandon explicit phoneme alignment and duration predictor. 
However, the implicit speech-text alignment and randomness of the sampling process tend to induce hallucination artifacts in synthesized speech, resulting in mispronunciations, audile noise, and abnormal pauses (as shown in Figure \ref{fig:error}). 

To mitigate pronunciation errors in synthesized speech, previous studies have deployed a two-stage pipeline to address this issue. The first stage recognizes the specific text associated with pronunciation error and the corresponding range in speech. The second stage is based on a text-based speech editing process that regenerates mispronounced speech segments ~\cite{bae2025unitcorrect, peng2024voicecraft}. Although these models indeed improve the fidelity and intelligence of generated speech, they depend on a complex apparatus of external tools (e.g., ASR models, SSL Models, MFA) to identify mispronunciations, and they cannot locate the segments with low audio quality. 

Moreover, inspired by the successful application of reinforcement learning from human feedback (RLHF) ~\cite{christiano2017deep} in calibrating the output of LLMs to better align with human preferences ~\cite{ouyang2022training}, recent studies explore different preference alignment methods to improve the intelligibility of zero-shot TTS models ~\cite{zhang2024speechalign,chen2024enhancing,hu2024robust,tian2025preference}. In particular, the preference alignment is performed by maximizing the rewards from diverse feedback to align the TTS model with human preference. However, collecting fine-grained preference rewards is challenging, hampered by costly annotations and complex pipelines that incorporate pre-trained assessment tools ~\cite{yao2025fine}.

To address the above issues, we present a unified multi-level Vox-Evaluator to facilitate correction of erroneous speech segments and guide preference alignment. The Vox-Evaluator is capable of identifying the temporal scope of speech segment with pronunciation errors or quality issues, detecting mismatch text, and evaluating the overall quality level of the synthesized speech. Based on the detection information obtained from Vox-Evaluator, we develop a fine-grained speech correction to regenerate speech segments with mispronunciations or audio quality issues. 
In addition, we demonstrate that the Vox-Evaluator can also serve as a fine-grained reward model to guide preference optimization on zero-shot TTS system.

In conclusion, our contributions are as follows: 
\begin{itemize}
\item We introduce a novel multi-level Vox-Evaluator that provides a comprehensive evaluation of the generated speech for the correction process and fine-grained guidance for preference alignment. The information contains the location of specific segments, detection of content and quality evaluation for the synthesized speech.
\item Based on Vox-Evaluator, we utilizes speech correction mechanism to enhance the stability and provide effective guidance for preference alignment on zero-shot TTS system, boosting the overall performance of the TTS model.
\item Experimental results on different zero-shot TTS frameworks prove that the Vox-Evaluator effectively facilitate the stability and fidelity of the TTS model.
\end{itemize}

\begin{figure*}[!t]
    \centering
    \includegraphics[width=0.84\textwidth]{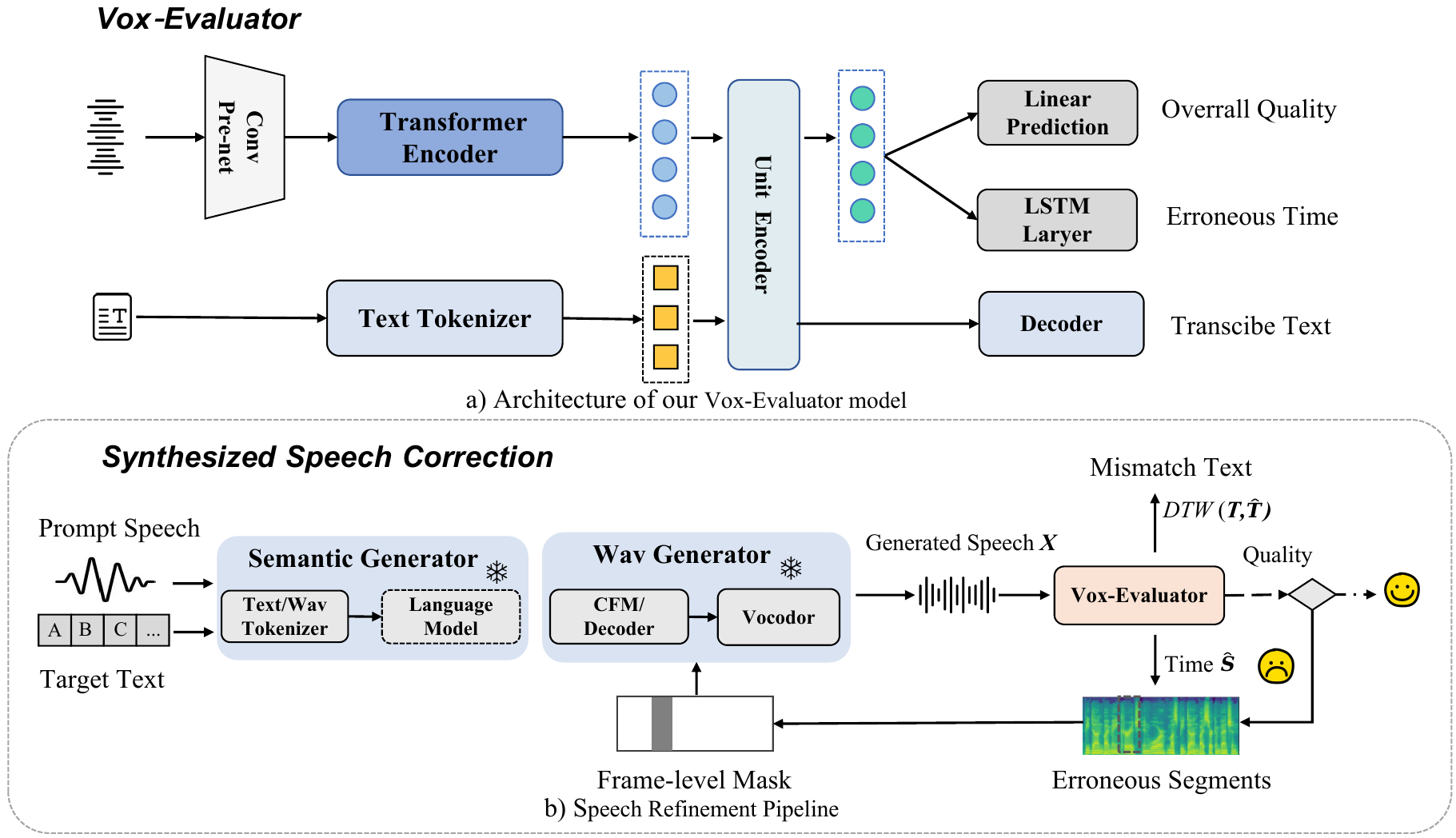}
\caption{a) Architectural overview diagram of the proposed Vox-Evaluator. b) The illustration of erroneous speech correction based on editing, the dashed box indicates that this module is optional.}
    \label{fig:model}
\end{figure*}

\section{Related Work}

\subsection{Speech Editing}
The speech editing which aims to alter specific words or phrases in the audio and keep the other regions unchanged. The traditional cut-copy-paste method ~\cite{morrison2021context} involves a simple process of cutting and pasting audio segments, but it may lead to prosody mismatch or boundary artifacts. In recent years, text-based speech editing systems have seen significant development, enabling users to modify an audio waveform by simply editing its transcript. Previous systems, including CampNet ~\cite{wang2022context} and FluentSpeech ~\cite{borsos2022speechpainter}, perform editing according to a masked reconstruction principle. They regenerate the target portion based on its surrounding acoustic context. While in the unified model, Voicebox ~\cite{le2023voicebox} provides a versatile framework based on flow matching to address both speech continuation and editing. VoiceCraft ~\cite{peng2024voicecraft} relies on an AR model to predict multi-layer acoustic tokens, enabling it to perform long-form text editing. Speech editing is the crucial part of the overall erroneous speech correction task. The broader refinement process involves other stages, such as automatic error detection and quality evaluation.

\subsection{Preference Alignment for TTS}
Recent works have performed preference alignment on zero-shot TTS models to enhance the overall system performance, including intelligibility, speaker similarity, emotional controllability and others.
For intelligibility, previous studies ~\cite{zhang2024speechalign,chen2024enhancing} directly leveraged WER as a direct reward signal or guiding metric to construct preference pairs for preference alignment.
For speaker similarity, some works ~\cite{sun2025f5r, hussain2025koel} utilize speaker verification score between the embeddings of the prompt audio and the generated audio as the reward to conduct preference alignment. 
For emotion controllability, emotional preferences and human-feedback have been introduced to improve emotional expressiveness and controllable style rendering ~\cite{gao2025emo, chen2024enhancing}.
Nevertheless, the potential of employing a fine-grained, multi-level reward model for preference alignment receives limited attention in the context of preference optimization on zero-shot TTS systems.

\section{Methodology}
In this section, we present Vox-Evaluator, a multi-level evaluator that can identify the temporal location of errors and evaluate the overall quality of synthesized speech.
Guided by the Vox-Evaluator, we first automatically locate faulty segments and then regenerate them in the speech refined process. To further improve the stability of the TTS model, the assessments of the evaluator are leveraged as a preference signal to efficiently guide fine-grained preference alignment. Moreover, for training the Vox-Evaluator, we introduce the FGES (fine-grained erroneous speech) dataset, a diverse dataset providing erroneous information and quality level.

\subsection{Architecture of the Vox-Evaluator}
To overcome the computational burden of large automatic speech recognition (ASR) models and complex dynamic time warping (DTW) and Montreal Forced Aligner (MFA), we introduce a unified, multi-level Vox-Evaluator. The evaluator aims to predict the timestamps of faulty speech segments, detect semantic content, and predict a holistic quality score for the entire speech sample. Built on previous speech-text alignment works, such as SpeechLM ~\cite{zhang2024speechlm} and STFT ~\cite{tang2022unified}, our evaluator is designed as an encoder-decoder architecture, which is shown in Figure~\ref{fig:model}a, comprising a speech encoder, phoneme tokenizer, unit encoder, timestamp predictor, overall quality score predictor, and a text decoder.

\begin{itemize}
\item \textbf{Speech Encoder}
A feature extraction module consisting of 1-D convolutional layers converts the synthesized speech into a sequence of hidden features that serve as input to the speech encoder. This speech encoder adopts a transformer architecture ~\cite{vaswani2017attention} based on self-supervised wav2vec2.0 ~\cite{baevski2020wav2vec} to model contextual information within the hidden feature sequence. 

\item \textbf {Unit Encoder}
The unit encoder adopts the same architecture as the encoder of the pre-trained language model BART ~\cite{lewis2019bart}. The unit encoder receives concatenated dual-modal inputs, including speech semantic tokens extracted from the speech encoder and target text tokens obtained from a phoneme tokenizer. This unit encoder facilitates mutual attention between the two modality features through bidirectional self-attention.

\item \textbf {Text Decoder}
The text decoder is the Transformer architecture inheriting from BART decoder, comprising a text embedding layer, multiple stacked Transformer layers, and a final output layer. The decoder's objective is to generate the target text sequence in the autoregressive manner, conditioned on the representations from the unit encoder. Discrepancies such as missing or mismatched text content are detected by comparing the generated sequence with the ground-truth text.

\item \textbf {Predictor}
We employ a multi-layer perceptron (MLP) to predict the overall quality of the input synthesized speech. The MLP consists of three linear layers, with each layer followed by layer normalization and GeLU activation ~\cite{hendrycks2016gaussian}. A stack of two long-short-term memory (LSTM) layers, followed by a linear layer and a sigmoid activation, is used to predict the timestamps of the speech segments with errors. The hidden features corresponding to the speech modality in the unit encoder's output are fed to the two predictors. Since the hidden features of speech have already addressed the information of the target text through the unit encoder, this facilitates the timestamp predictor to detect segments of synthesized speech that exhibit semantic mismatches with the target text.

\end{itemize}
\subsubsection{Training Strategy}
We employ the well-trained checkpoint from speech-text model ~\cite{tang2022unified} which is pre-trained with self-supervised speech masked token prediction, supervised speech phoneme alignment, and a supervised speech-to-text task. 
To improve the model's diverse capabilities, we employ a multi-task learning framework where each prediction head is trained with a specialized objective. To address the severe class imbalance in timestamp prediction, we utilize a frame-wise focal loss ~\cite{lin2017focal}. The text token prediction is trained with a standard token-level cross-entropy loss, while the speech quality score prediction is optimized using a Mean Squared Error (MSE) loss.
The overall training objective is a composite loss combining frame-wise focal loss, MSE loss, and token-level cross-entropy loss. The loss function is formulated as follows:
\begin{align}
\mathbf{L}_{\text{total}} = L_{\text{mse}} + L_{\text{frame}} + L_{\text{ce}}
\end{align}

\subsection{Synthesized Speech Correction}
\paragraph{Error Detection and Correction}
As shown in Figure~\ref{fig:model}b, the error detection is a dual task that includes both locating the specific speech segments with pronunciation errors and verifying the text content against the ground-truth. Given the prompt and target text $T$, a semantic generator will produce semantic representations which are passed to a wav generator for synthesizing speech $X$. Our multi-level Vox-Evaluator is then employed to predict the corresponding text sequences $\hat{T}$ and the time scope $\hat{S}$ associated with erroneous segments in the synthesized speech. Discrepancies between the reference text and the predicted text frequently occur in challenging cases. Consequently, it is necessary to perform error identification by comparing the predicted text $\hat{T}$ with the ground-truth prompt $T$. 

We formulate locating semantic inconsistencies as an optimization problem using Dynamic Time Warping (DTW). Specifically, we attempt to find a warping path $\pi$ that minimizes the cumulative cost between the word sequence $T=(t_1,t_2,\dots,t_n)$ and $\hat{T}=(\hat{t}_1,\hat{t}_2,\dots,\hat{t}_m)$. The optimization objective is to find the optimal path $\pi^{\ast}$ that minimizes the total alignment cost:
\begin{align}
\pi^{\ast}
= \arg\min_{\pi}
\sum_{(i,j)\in\pi} C\!\left(t_i,\hat{t}_j\right),
\end{align}
where $\pi=((i_1,j_1),\dots,(i_K,j_K))$ is a sequence of index pairs $(i,j)$ defining the alignment, and $C\!\left(t_i,\hat{t}_j\right)$ is a cost function that measures the dissimilarity between the ground-truth word $t_i$ and the recognized word $\hat{t}_j$ (e.g., based on phonetic distance). The resulting alignment path provides a detailed mapping of discrepancies between the intended and recognized text. 

To address the erroneous segments, we utilize a mask when a non-empty time scope is predicted by Vox-Evaluator or the DTW alignment reveals discrepancies. With the $\hat{S}$ time scope, we then create a speech mask that divides the sequence into parts to be corrected and parts to be preserved. To account for potential inaccuracies in the initial time detection, we apply a small margin to extend temporal masks. This margin is derived by uniformly partitioning the duration of the mismatched text, which guarantees the complete localization and removal of erroneous segments.
We refine segment-level speech errors with an editing TTS model that leverage the generated segmentation masks in the speech representation and the text prompt as conditions for the speech generation process, resulting in significantly lower computational costs compared to complete generation.
This process is repeated for two iterations to effectively enhance the generated stability. We empirically discuss this in the ablation study. The speech correction can be summarized in Algorithm~\ref{alg:sc}. 

\paragraph{Evaluation Mechanism}
To judge whether synthesized speech requires error correction, we conduct a comprehensive evaluation to select low-fidelity synthesized samples for refinement. Our evaluation methodology considers two key aspects: (i) an overall audio quality score, predicted by a quality assessment module, and (ii) a word error rate, which quantifies the semantic divergence from the text decoder. By incorporating these two complementary aspects into our evaluation, we ensure that the refined synthesized speech maintains both acoustic quality and semantic integrity. Finally, through this process of continuous refinement, the stability and intelligibility of zero-shot TTS systems are significantly enhanced, without any additional model retraining or fine-tuning.

\begin{algorithm}[t]
\caption{A pseudo-code of speech correction.}
\label{alg:sc}
\begin{algorithmic}[1]
\Function{Correction}{Speech $X$, Text $T$, Time scope $\hat{S}$, Maximum iterations $max\_iter$}
  \While{$t < max\_iter$}
    \State $\hat{S}, \hat{T} \gets \Call{Vox-Evaluator}{X, T}$
    \State $\mathcal{O} \gets \Call{DTW}{T, \hat{T}}$ \Comment{Find mismatch errors}
    \State $X_{\text{edit}} \gets \Call{Refine}{X, \hat{S}, \mathcal{O}}$
    \If{$\mathcal{O} \neq \varnothing$}
      \State $T, \hat{S} \gets \Call{Update}{T, \hat{S}, \mathcal{O}}$
    \Else
      \textbf{ Break}
    \EndIf
    \State $t \gets t + 1$
    \EndWhile
    \State \Return $X_{\text{edit}}$
\EndFunction
\end{algorithmic}
\end{algorithm}

\subsection{Vox-Evaluator Guided Fine-grained Preference Alignment}
To achieve comprehensive improvements in speech generation, we employ the Vox-Evaluator to evaluate the speech quality from semantic correctness and audio quality perspectives. 
The work in ~\cite{zhang2025advancing} introduces the utilization of preference pairs towards direct preference learning. 
We select two speech samples generated under the same conditions from our generated datasets. The sample exhibiting higher fidelity or quality is designated as the ``winning'' or preferred sample, denoted as $x^w$, while the sample containing erroneous parts or demonstrating lower quality is designated as the ``losing'' sample, $x^l$. The objective is as follows: 
\begin{multline}
L(\theta) = -\mathbb{E}_{\substack{
    (x^w_0, x^l_0) \sim \mathcal{D}, \, t \sim \mathcal{U}(0,T) \\
    x^w_t \sim q(x^w | x^w_0), \, x^l_t \sim q(x^l | x^l_0)
}}
\Biggl[ \log \sigma \Biggl( -\beta/2 \\
\Big( ( \| \varepsilon^w - \varepsilon_\theta (x^w_t, t) \|^2 - \| \varepsilon^w - \varepsilon_{\text{ref}} (x^w_t, t) \|^2 ) \\
- ( \| \varepsilon^l - \varepsilon_\theta (x^l_t, t) \|^2 - \| \varepsilon^l - \varepsilon_{\text{ref}} (x^l_t, t) \|^2 ) \Big) \Biggr) \Biggr]
\end{multline}
where \( t \sim \mathcal{U}(0, T) \) and \( x_t \sim q(x_t | x_0) \), \( \beta \) controls the degree of divergence.

Although TTS-DPO has designed a direct preference optimization on TTS models, it suffers from information redundancy and over-optimization on the full utterance. To address this, we propose a targeted strategy to ensure a more fine-grained enhancement of the generated speech. Specifically, instead of calculating the loss for the entire audio, we use timestamp annotations to focus on specific segments and compute the attentive loss only for those segments.

\section{Dataset for Vox-Evaluator}
\subsection{Data Construction}
To train the Vox-Evaluator, we construct a dataset consisting of synthesized speech samples with various issues. These issues include: (1) common issue consisting of mispronunciations and omissions, (2) repeated issue consisting of repetitive and redundant pronunciation, and (3) punctuation issue consisting of unnatural pause and prosody. Specially, we first selected a subset text prompts from the Emilia-Large corpus ~\cite{he2024emilia} which provides a rich source of real-world speech data across diverse topics and styles. Then we randomly sample speech prompts from LibriTTS ~\cite{zen2019libritts} and use the pretrained TTS model ~\cite{chen2024f5,peng2024voicecraft} to synthesize audio samples with varying temperature and guidance scale. 
Besides, we utilize data augmentation to produce low-quality samples with noise or abnormal pauses, which we classify as (4) abnormal issues shown in Table~\ref{tab:datasets}. The motivation behind this augmentation is to intentionally infuse the dataset with corrupted speech. This dataset enhances the model's audio quality perception capability during training and facilitates the evaluation of its performance in detecting erroneous segments in synthesized speech.

Our annotation pipeline involves three distinct stages to label semantic content, timestamps of problematic pronunciations, and audio quality levels. First, we employ the Whisper-Large-v3 model to transcribe the synthesized speech and then compare the transcripts against the target text. Second, for fine-grained timestamp labeling, we utilize the Montreal Forced Aligner (MFA) ~\cite{kurzinger2020ctc} to locate the word-level pronunciation and ensure that the duration of word-level alignments matches the corresponding frame-level speech representations. Finally, for audio quality assessment, we perform sentence-level score annotation on the augmented synthesized speech using the Audiobox-Aesthetics ~\cite{tjandra2025meta} and manual annotation. The overall speech quality annotation is a composite score that combines the average of the predicted content enjoyment score and the production quality score. The resulting score ranges from 1 to 10, representing an audio quality metric consistent with human aesthetic perception.

\subsection{Data Summarization}
We designate the constructed synthesized speech corpus as FGES dataset, which consists of 22k speech samples annotated with timestamps of error segments and quality level. The dataset is divided into a training set of 20k samples, a validation set of 1k samples, and a test set of 1k samples. To ensure a balanced distribution, we specifically selected samples that cover a wide range of error types. Furthermore, we conducted a subjective evaluation to verify that the dataset's annotations align with human perception of intelligibility.
Table ~\ref{tab:datasets} shows the statistics for each issue. 

\begin{table}[!t]
    \centering
    \small
    \begin{tabular}{lccc}
    \toprule
    Types & \# Size & \# Avg. words & \# Avg. audio len \\
    \midrule
    Common issue & 12.0K & 36 & 19.3s \\
    Repeated issue & 3.9K & 27 & 15.8s \\
    Punctuation issue & 3.5K  & 21 & 10.7s \\
    Abnormal issue & 1.6K & 24 & 12.5s \\
    \bottomrule
\end{tabular}
\caption{Details of the FGES dataset.}
\label{tab:datasets}
\end{table}

\begin{table*}[t]
\centering
\small
\renewcommand{\arraystretch}{1.2}
\setlength{\tabcolsep}{3.5pt}
\begin{tabular}{lcccccc} 
\toprule
\multirow{2}{*}{\textbf{Model}} & \multirow{2}{*}{\textbf{Params}} & \multicolumn{2}{c}{\textbf{Time Scope}} & \multicolumn{2}{c}{\textbf{Quality Score}} & \multicolumn{1}{c}{\textbf{Text Accuracy}} \\
\cmidrule(lr){3-4} \cmidrule(lr){5-6} \cmidrule(lr){7-7} 
& & {MSE (↓)} & {IOU (↑)} & {utt-PCC (↑)} & {sys-SRCC (↑)} & {WER (\%↓)} \\
\midrule
SenseVoice & 234M & - & - & - & - & 3.43 \\
Whisper-S & 224M & - & - & - & - & 3.05 \\
Wav2vec2.0 (fine-tuned) & 220M & 0.0049 & 0.653 & 0.459 & 0.463 & 2.96 \\
\midrule
Vox-Evaluator wo/pretrained & 185M & 0.0058 & 0.435 & 0.398 & 0.516 & 3.15 \\
Vox-Evaluator & 185M & \textbf{0.0028} & \textbf{0.782} & \textbf{0.541} & \textbf{0.630} & \textbf{2.64} \\
\bottomrule
\end{tabular}
\caption{Fine-grained prediction results on the FGES test set. `wo/pretrained' denotes the Vox-Evaluator without pre-training.}
\label{tab:prediction}
\end{table*}

\section{Experiments}
\subsection{Experimental Setups} 
\paragraph{Evaluation Dataset}
For speech correction, we employ Seed-TTS test-en ~\cite{anastassiou2024seed} and LibriSpeech-PC ~\cite{meister2023librispeech} test-clean as test data. For ablation, we incorporate TTSDS2 ~\cite{minixhofer2025ttsds2} to evaluate system stability and audio quality. 

\paragraph{Implementation Details}
The speech encoder of the Vox-Evaluator is composed of a 6-layer transformer with a 768-dimensional hidden state, a 3072-dimensional FFN, and 8 attention heads. For temporal error prediction, we employ a timestamp predictor consisting of 2 bidirectional LSTM layers and a final linear layer with sigmoid activation. The quality score predictor is an MLP with three dense layers (output sizes 768, 768, and 1), using Layer Normalization and GeLU activations. The entire model is fine-tuned for 50 epochs with batch size of 24 on our training dataset using the Adam optimizer and a learning rate of 1e-4. The long waveform is divided into small segments of 30 seconds.

\paragraph{Evaluation Metrics} 
\begin{itemize}
    \item \textbf{Metrics for Vox-Evaluator:} We adopt two standard metrics to measure the performance of Vox-Evaluator in predicting overall audio quality: the utterance-level Pearson linear correlation coefficient (utt-PCC) ~\cite{benesty2009pearson} and the system-level Spearman’s rank correlation coefficient (sys-SRCC) ~\cite{sedgwick2014spearman}. For the localization of erroneous segments, we evaluate Vox-Evaluator using Intersection over Union (IOU). Additionally, to assess the model’s performance in identifying the absence of errors, we compute the Mean Squared Error on all well-pronounced speech samples. For assessing the transcription capability, we adopt word-level Error Rate (WER) as the measure metric. Specifically, these metrics are calculated across all test samples. 
    \item \textbf{Metrics for TTS Model:} To measure the performance of preference alignment and speech correction, we report the WER for objective evaluation, using Whisper-large-v3 ~\cite{radford2023robust} for transcription. Speaker similarity is also assessed via the cosine similarity of speaker embeddings extracted by a WavLM-large-based model ~\cite{chen2022wavlm}. For subjective evaluation, we measure naturalness using Comparative Mean Opinion Scores (CMOS), where human evaluators compare the synthesized speech to the ground truth. And we also assess the naturalness of the synthesized speech using the UTokyo-sarulab mean opinion score (UTMOS) prediction system ~\cite{saeki2022utmos}, which serves as an automatic and efficient metric of speech quality.
\end{itemize}

\paragraph{Benchmark Methods} 
Since Vox-Evaluator is a multi-level evaluator, we employ a fine-tuned wav2vec2.0 model trained on our FGES dataset as a baseline. To assess its accuracy in transcription, we directly compare our Vox-Evaluator with high-performing ASR models, including SenseVoice and Whisper. In addition, we select F5-TTS and VoiceCraft for error correction experiments, since they are specifically designed for this purpose. We also employ several prominent zero-shot TTS models as baselines, including CosyVoice ~\cite{du2024cosyvoice}, NaturalSpeech 3 ~\cite{ju2024naturalspeech}, MaskGCT ~\cite{wang2024maskgct}, and Llasa ~\cite{ye2025llasa}.

\begin{table}[t]
\centering
\small
\renewcommand{\arraystretch}{1.2}
\setlength{\tabcolsep}{1.0pt}
\begin{tabular}{lccc}
\hline
\toprule
\textbf{Model} & \textbf{WER(\%↓)} & \textbf{SIM-o}(↑) & \textbf{CMOS(↑)} \\ \hline
\multicolumn{4}{c}{\textit{Seed-TTS test-en}} \\ \hline
CosyVoice\cite{du2024cosyvoice} & 4.08 & 0.64 & 0.02 \\ 
NaturalSpeech 3\cite{ju2024naturalspeech}  & 1.94 & 0.67 & 0.16 \\ 
MaskGCT\cite{wang2024maskgct} & 2.01 & 0.69 & 0.12 \\
Llasa-1B\cite{ye2025llasa} & 2.03 & 0.76 & 0.23 \\ 
VoiceCraft\cite{peng2024voicecraft} & 7.56 & 0.47 & -1.08 \\ 
F5-TTS\cite{chen2024f5} & 1.73 & 0.67 & 0.31 \\ \hline
VoiceCraft$_{refine}$ & \textbf{5.11} & 0.47 & -0.78 \\ 
F5-TTS$_{refine}$ & \textbf{1.42} & \textbf{0.68} & \textbf{0.33} \\ \hline
\multicolumn{4}{c}{\textit{LibriSpeech-PC test-clean}} \\ \hline
VoiceBox\cite{le2023voicebox} & 2.03 & 0.64 & -0.41 \\ 
MaskGCT & 2.63 & 0.69 & 0.13 \\
VoiceCraft & 4.68 & 0.45 & -0.33 \\ 
F5-TTS & 2.42 & 0.66 & -0.22 \\ \hline
VoiceCraft$_{refine}$ & \textbf{3.16} & 0.44 & -0.26 \\ 
F5-TTS$_{refine}$ & \textbf{2.03} & \textbf{0.66} & \textbf{-0.18} \\ 
\bottomrule 
\end{tabular}
\caption{Performance comparison of speech correction with other systems on Seed-TTS test-en. $refine$ denotes the proposed correction process.}
\label{tab:refine}
\end{table}
\subsection{Main Results}
\paragraph{Performance of the Vox-Evaluator}
Table~\ref{tab:prediction} presents the performance comparison of Vox-Evaluator and baselines in different metrics. Regarding timestamp prediction, we first report MSE for test samples containing no erroneous segments. It evaluates the performance on correct samples that are free of any artifacts. For the mispronounced samples, the Vox-Evaluator achieves a high IOU of 0.782, surpassing the fine-tuned wav2vec2.0 baseline by a substantial margin (+19.7\% gains). This result demonstrates that the Vox-Evaluator effectively models the alignment between the speech semantic representation and the ground-truth text content. 
The performance of wo/pre-trained suggests that using a pre-trained checkpoint is crucial, as the fine-grained metric drops significantly when it is randomly initialized. 
In terms of quality score prediction, the Vox-Evaluator achieves utt-PCC and sys-SRCC scores of 0.541 and 0.630, respectively, demonstrating the significant positive correlation between the annotated scores and the predicted scores.
For the text accuracy, we compare our Vox-Evaluator with several ASR models on the task of speech transcription. The results show that our model achieves a superior Word Error Rate (WER) of 2.64\% with fewer parameters, outperforming other methods such as Whisper and SenseVoice. It is particularly remarkable that the Vox-Evaluator can perceive cross-modal information between text and speech, which directly boosts its effectiveness in fine-grained tasks.
\begin{figure}[!t]
    \centering
    \includegraphics[width=1.\columnwidth]{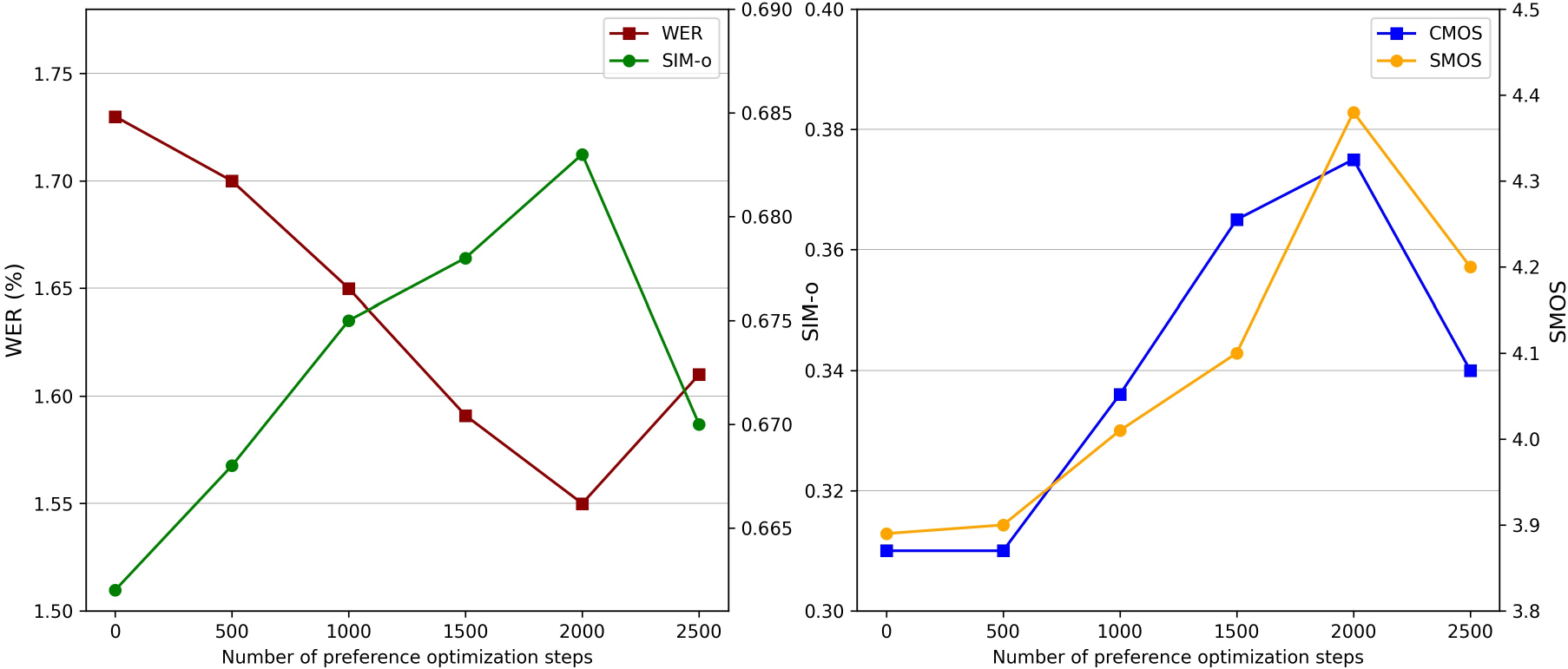}
    \caption{Performance of F5-TTS model with fine-grained preference alignment on the Seed-TTS test set at different training steps.}
    \label{fig:dpo}
\end{figure}

\begin{figure}[!t]
    \centering
    \includegraphics[width=1.\columnwidth]{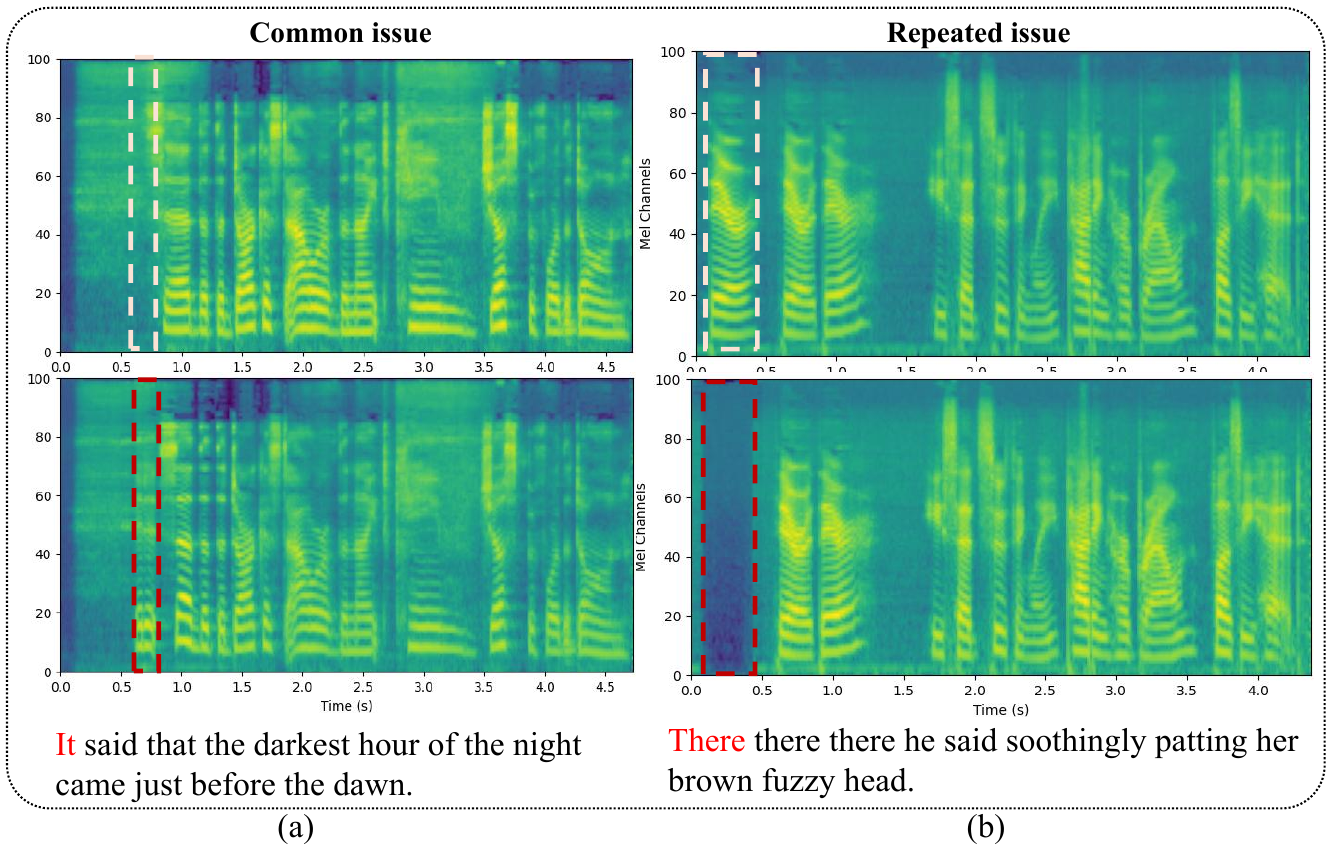}
    \caption{Examples of corrected speech through speech correction.}
    \label{fig:case}
\end{figure}

\paragraph{Performance of the Speech Correction}
While the zero-shot TTS models are prone to generate low-quality or erroneous content for hard cases, the speech correction yields outputs demonstrating both semantic correctness and natural prosodic coherence.
As shown in Table~\ref{tab:refine}, we validate the effectiveness of the speech correction mechanism on the Seed-TTS test-en and Librispeech-PC test-clean datasets. When we compare the performance with different baselines, we observe that NAR-based refinement models perform better than most models, achieving enhanced stability. 
Besides, compared to the original VoiceCraft, which is AR-based model, the correction process also improves the performance significantly. 
In general, F5-TTS$_{refine}$ and VoiceCraft$_{refine}$ obtain significant reductions in WER compared to the Common baseline by 21\%, 32\% respectively. The refined models demonstrate consistently strong performance and robustness across different datasets. These improvements can be attributed to Vox-Evaluator's multi-level capability.

Figure~\ref{fig:case} shows several examples in speech correction, demonstrating how Vox-Evaluator rectifies speech segments with errors. The first column demonstrates a case where the speech correction process accurately recovers missing semantic information, thereby enhancing the word-level duration alignment between the speech and text. The second column presents an instance where we successfully removes an unnecessary or erroneous speech segment. 

\paragraph{Fine-grained Preference Alignment}
Zero-shot TTS model benefits significantly from fine-grained multi-level preference alignment through Direct Preference Optimization (DPO). As illustrated in Figure~\ref{fig:dpo}, with increasing iteration steps, the performance of F5-TTS shows a progressive improvement over the original model in terms of both subjective and objective evaluation metrics.
Notably, in terms of objective evaluation, the WER decreases from 1.73\% to 1.55\% and Sim-o increases to a higher score of 0.683. In subjective evaluation, the preference optimization demonstrates a continuous improvement in CMOS and SMOS scores throughout the 2000 training steps. These results demonstrate that fine-grained alignment with DPO not only enhances perceptual quality but also improves semantic coherence, leading to a more holistic and stable generative TTS model.
In the early training stage, the preference information facilitate substantial adjustments and enable rapid improvements. However, continued training leads to a decrease in model performance. This is possibly because the feedback reward becomes saturated when high-quality samples exhibit minimal variation, which increases the difficulty of preference optimization and may even degrade system performance.

\begin{figure}[!t]
    \centering
    \includegraphics[width=0.8\columnwidth]{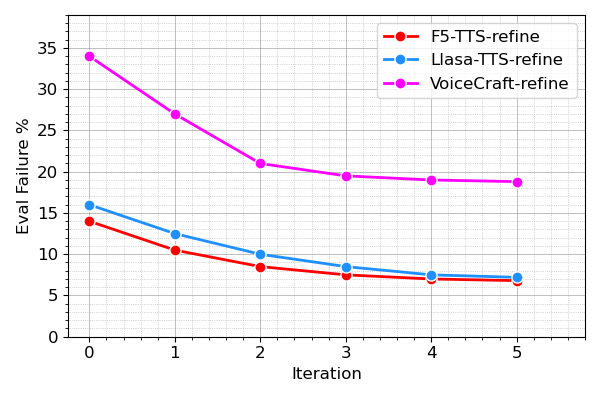}
    \caption{Iterative speech correction and resulting TTSDS2 evaluation failures.}
    \label{fig:pic}
\end{figure}

\begin{table}[!t]
\centering
\small
\renewcommand{\arraystretch}{1.2}
\setlength{\tabcolsep}{1.pt}
\begin{tabular}{lcccc}
\toprule
\textbf{Model} & \textbf{Error Detect} & \textbf{Quality Eval} & \textbf{Failure(\%↓)} & \textbf{UTMOS (↑)} \\
\midrule
\multirow{4}{*}{VoiceCraft} 
&            &            & 34.0  & 2.61 \\ 
& \checkmark &            & 22.0 & 2.80\\
&            & \checkmark & 30.0 & 2.82\\
& \checkmark & \checkmark & \textbf{20.0} & \textbf{2.89}\\ 
\cmidrule(lr){1-5}
\multirow{4}{*}{F5-TTS} 
&            &            & 14.0  & 3.35 \\
& \checkmark &            & 6.0 & 3.59\\
&            & \checkmark & 10.0 & 3.66\\
& \checkmark & \checkmark & \textbf{6.0} & \textbf{3.70}\\
\bottomrule
\end{tabular}
\caption{Ablation study for speech correction process on TTSDS2.}
\label{tab:ablidation}
\end{table}

\subsection{Ablation Study}
To demonstrate the effectiveness of Vox-Evaluator in guiding speech correction, we conducted ablation experiments on our proposed method. 
The first experiment employed error detection solely to identify erroneous segments and mismatched text.
The second experiment employed only the quality evaluation method to assess the synthesized speech.
The third combined both the error detection and quality evaluation. We employ the failure rate introduced by TTSDS2 to evaluate stability and UTMOS to evaluate speech quality.
Table \ref{tab:ablidation} shows that both error detection and quality evaluation can achieve performance improvement over the baseline, and error detection has a more significant impact. The combination of both achieved the best performance, though only slightly better than error detection alone. Notably, error detection is highly effective at reducing mispronunciations, reducing the failure rate of the F5-TTS model from 12\% to 6\%. Furthermore, a comparison of results on the quality evaluation suggests that F5-TTS is less prone to speech corruption than VoiceCraft.

We also illustrate the failure rate declining over iterations in Figure \ref{fig:pic} for several backend models. Our speech correction approach achieves a consistent reduction in failures as iterations increase. However, the improvement begins to plateau after two iterations. Therefore, we set the number of iterations to two.

\section{Conclusion}
We propose a multi-level Vox-Evaluator to enhance stability and fidelity through speech correction mechanism and fine-grained preference alignment. The Vox-Evaluator facilitates the identification of erroneous speech segments and mismatched text tokens, providing guidance in the correction process. Moreover, leveraging the Vox-Evaluator, the speech correction process corrects erroneous segments in the speech generated by zero-shot TTS models and preserves the overall quality of the speech.
We show that using Vox-Evaluator can further guide preference alignment to enhance the performance of the TTS system. Extensive experiments demonstrate the effectiveness of the Vox-Evaluator in providing fine-grained reward. Our approach does not require fine-tuning generative model and is plug-gable to other existing TTS models.
\bibliography{aaai2026}

\end{document}